\pdfoutput=1
\documentclass[12pt,a4paper]{article}
\usepackage{ifthen} % for conditional statements
\newboolean{pdflatex}
\setboolean{pdflatex}{true} % False for eps figures 

\usepackage{tikzscale}
\usepackage{tikz-feynman}

\usepackage{enumitem}

\newboolean{articletitles}
\setboolean{articletitles}{true} % False removes titles in references

\newboolean{uprightparticles}
\setboolean{uprightparticles}{false} %True for upright particle symbols

\def\paperauthors{Chen Chen, Tim Gershon, Thomas Latham} % Leave as is for PAPER, CONF and FIGURE
\def\paperasciititle{Method to study CP violation in Bs -> KS K pi decays} % Set ASCII title here !! MAKE sure it's only ASCII characters !! 
\def\papertitle{Method to study \CP violation in \BsToKSKPi decays} % Latex formatted title
\def\paperkeywords{{High Energy Physics}} % Comma separated list
\def\papercopyright{\the\year\ CERN for the benefit of the LHCb collaboration} % new since 9/Apr/2018

\def\paperlicenceurl{https://creativecommons.org/licenses/by/4.0/}

\newif\ifEnableSectionTOCLinks
\EnableSectionTOCLinksfalse % deactivated

\usepackage[top=1in, bottom=1.25in, left=1in, right=1in]{geometry}

\usepackage{microtype}
\usepackage{lineno}  % for line numbering during review
\usepackage{xspace} % To avoid problems with missing or double spaces after
\usepackage{caption} %these three command get the figure and table captions automatically small

\usepackage{graphicx}  % to include figures (can also use other packages)
\usepackage{color}
\usepackage{colortbl}
\graphicspath{{./}} % Make Latex search root dir for figures
\usepackage{amsmath} % Adds a large collection of math symbols
\usepackage{amssymb}
\usepackage{amsfonts}
\usepackage{upgreek} % Adds in support for greek letters in roman typeset

\newcommand*\patchAmsMathEnvironmentForLineno[1]{%
\expandafter\let\csname old#1\expandafter\endcsname\csname #1\endcsname
\expandafter\let\csname oldend#1\expandafter\endcsname\csname
end#1\endcsname
 \renewenvironment{#1}%
   {\linenomath\csname old#1\endcsname}%
   {\csname oldend#1\endcsname\endlinenomath}%
}
\newcommand*\patchBothAmsMathEnvironmentsForLineno[1]{%
  \patchAmsMathEnvironmentForLineno{#1}%
  \patchAmsMathEnvironmentForLineno{#1*}%
}
\AtBeginDocument{%
\patchBothAmsMathEnvironmentsForLineno{equation}%
\patchBothAmsMathEnvironmentsForLineno{align}%
\patchBothAmsMathEnvironmentsForLineno{flalign}%
\patchBothAmsMathEnvironmentsForLineno{alignat}%
\patchBothAmsMathEnvironmentsForLineno{gather}%
\patchBothAmsMathEnvironmentsForLineno{multline}%
\patchBothAmsMathEnvironmentsForLineno{eqnarray}%
}

\usepackage[pdftex,
            pdfauthor={\paperauthors},
            pdftitle={\paperasciititle},
            pdfkeywords={\paperkeywords}]{hyperref}
\usepackage{hyperxmp}
\hypersetup{
    pdfcopyright={Copyright (C) \papercopyright},
    pdflicenseurl={\paperlicenceurl}
}
\usepackage[bottom,flushmargin,hang,multiple]{footmisc}

\usepackage[all]{hypcap} % Internal hyperlinks to floats.

\usepackage{xspace} 
\usepackage{upgreek}

\def\MagUp {\mbox{\em Mag\kern -0.05em Up}\xspace}

\ifthenelse{\boolean{uprightparticles}}%
{

 \def\Ppi         {\ensuremath{\uppi}\xspace}

 \def\Ppsi        {\ensuremath{\uppsi}\xspace}

 \def\PDelta      {\ensuremath{\Delta}\xspace}                 
 \def\PXi         {\ensuremath{\Xi}\xspace}                 
 \def\PLambda     {\ensuremath{\Lambda}\xspace}                 
 \def\PSigma      {\ensuremath{\Sigma}\xspace}                 
 \def\POmega      {\ensuremath{\Omega}\xspace}                 
 \def\PUpsilon    {\ensuremath{\Upsilon}\xspace}
 \let\oldPi\Pi
 \def\PPi         {\ensuremath{\oldPi}\xspace}

 \def\PB      {\ensuremath{\mathrm{B}}\xspace}                 
 \def\PD      {\ensuremath{\mathrm{D}}\xspace}                 
 \def\PJ      {\ensuremath{\mathrm{J}}\xspace}                 
 \def\PK      {\ensuremath{\mathrm{K}}\xspace}                 
 \def\Ps      {\ensuremath{\mathrm{s}}\xspace}

 \def\thebaroffset{0.0em}
}
{

 \def\Ppi         {\ensuremath{\pi}\xspace}

 \def\Ppsi        {\ensuremath{\psi}\xspace}                 
                  
 \mathchardef\PDelta="7101
 \mathchardef\PXi="7104
 \mathchardef\PLambda="7103
 \mathchardef\PSigma="7106
 \mathchardef\POmega="710A
 \mathchardef\PUpsilon="7107
 \mathchardef\PPi="7105
 \def\PB      {\ensuremath{B}\xspace}                 
 \def\PD      {\ensuremath{D}\xspace}                 
 \def\PJ      {\ensuremath{J}\xspace}                 
 \def\PK      {\ensuremath{K}\xspace}                 
 \def\Ps      {\ensuremath{s}\xspace}

 \def\thebaroffset{0.18em}
}
\newcommand{\offsetoverline}[2][\thebaroffset]{\kern #1\overline{\kern -#1 #2}}%

\makeatletter
\ifcase \@ptsize \relax% 10pt
  \newcommand{\miniscule}{\@setfontsize\miniscule{4}{5}}% \tiny: 5/6
\or% 11pt
  \newcommand{\miniscule}{\@setfontsize\miniscule{5}{6}}% \tiny: 6/7
\or% 12pt
  \newcommand{\miniscule}{\@setfontsize\miniscule{5}{6}}% \tiny: 6/7
\fi
\makeatother

\DeclareRobustCommand{\optbar}[1]{\shortstack{{\miniscule (\rule[.5ex]{1.25em}{.18mm})}
  \\ [-.7ex] $#1$}}

\def\squark    {{\ensuremath{\Ps}}\xspace}

\def\pion   {{\ensuremath{\Ppi}}\xspace}
\def\piz    {{\ensuremath{\pion^0}}\xspace}
\def\pip    {{\ensuremath{\pion^+}}\xspace}
\def\pim    {{\ensuremath{\pion^-}}\xspace}
\def\pipm   {{\ensuremath{\pion^\pm}}\xspace}
\def\pimp   {{\ensuremath{\pion^\mp}}\xspace}

\def\kaon    {{\ensuremath{\PK}}\xspace}
\def\Kbar    {{\ensuremath{\offsetoverline{\PK}}}\xspace}

\def\KorKbar {\kern \thebaroffset\optbar{\kern -\thebaroffset \PK}{}\xspace}
\def\Kz      {{\ensuremath{\kaon^0}}\xspace}

\def\Kp      {{\ensuremath{\kaon^+}}\xspace}
\def\Km      {{\ensuremath{\kaon^-}}\xspace}
\def\Kpm     {{\ensuremath{\kaon^\pm}}\xspace}
\def\Kmp     {{\ensuremath{\kaon^\mp}}\xspace}
\def\KS      {{\ensuremath{\kaon^0_{\mathrm{S}}}}\xspace}

\def\Kstarz  {{\ensuremath{\kaon^{*0}}}\xspace}
\def\Kstarzb {{\ensuremath{\Kbar{}^{*0}}}\xspace}
\def\Kstar   {{\ensuremath{\kaon^*}}\xspace}

\def\Kstarp  {{\ensuremath{\kaon^{*+}}}\xspace}
\def\Kstarm  {{\ensuremath{\kaon^{*-}}}\xspace}
\def\Kstarpm {{\ensuremath{\kaon^{*\pm}}}\xspace}

\def\D       {{\ensuremath{\PD}}\xspace}

\def\DorDbar {\kern \thebaroffset\optbar{\kern -\thebaroffset \PD}\xspace}

\def\Dp      {{\ensuremath{\D^+}}\xspace}
\def\Dm      {{\ensuremath{\D^-}}\xspace}

\def\DpDm    {\ensuremath{\Dp {\kern -0.16em \Dm}}\xspace}

\def\B       {{\ensuremath{\PB}}\xspace}
\def\Bbar    {{\ensuremath{\offsetoverline{\PB}}}\xspace}

\def\BorBbar {\kern \thebaroffset\optbar{\kern -\thebaroffset \PB}\xspace}

\def\Bd      {{\ensuremath{\B^0}}\xspace}

\def\BdorBdbar {\kern \thebaroffset\optbar{\kern -\thebaroffset \Bd}\xspace}

\def\Bs      {{\ensuremath{\B^0_\squark}}\xspace}
\def\Bsb     {{\ensuremath{\Bbar{}^0_\squark}}\xspace}
\def\BsorBsbar {\kern \thebaroffset\optbar{\kern -\thebaroffset \Bs}\xspace}

\def\Bds     {{\ensuremath{\B_{(\squark)}^0}}\xspace}

\def\BdorBs  {\Bds}

\def\jpsi     {{\ensuremath{{\PJ\mskip -3mu/\mskip -2mu\Ppsi}}}\xspace}

\def\Y#1S{\ensuremath{\PUpsilon{(#1S)}}\xspace}

\def\LorLbar     {\kern \thebaroffset\optbar{\kern -\thebaroffset \PLambda}\xspace}

\newcommand{\decay}[2]{\ensuremath{\mathinner{#1\!\to #2}}\xspace}

\def\to                 {\ensuremath{\rightarrow}\xspace}

\def\CP                {{\ensuremath{C\!P}}\xspace}

\newcommand{\dms}{{\ensuremath{\Delta m_{\squark}}}\xspace}

\newcommand{\DGs}{{\ensuremath{\Delta\Gamma_{\squark}}}\xspace}

\def\AT#1     {\ensuremath{A_{\mathrm{T}}^{#1}}\xspace}           % 2

\def\C#1      {\ensuremath{\mathcal{C}_{#1}}\xspace}                       % 9
\def\Cp#1     {\ensuremath{\mathcal{C}_{#1}^{'}}\xspace}                    % 7
\def\Ceff#1   {\ensuremath{\mathcal{C}_{#1}^{\mathrm{(eff)}}}\xspace}        % 9  
\def\Cpeff#1  {\ensuremath{\mathcal{C}_{#1}^{'\mathrm{(eff)}}}\xspace}       % 7
\def\Ope#1    {\ensuremath{\mathcal{O}_{#1}}\xspace}                       % 2
\def\Opep#1   {\ensuremath{\mathcal{O}_{#1}^{'}}\xspace}                    % 7

\newcommand{\aunit}[1]{\ensuremath{\text{\,#1}}}       
\newcommand{\tev}{\aunit{Te\kern -0.1em V}\xspace}
\newcommand{\gev}{\aunit{Ge\kern -0.1em V}\xspace}
\newcommand{\mev}{\aunit{Me\kern -0.1em V}\xspace}
\newcommand{\kev}{\aunit{ke\kern -0.1em V}\xspace}
\newcommand{\ev}{\aunit{e\kern -0.1em V}\xspace}
 
\newcommand{\mevc}{\ensuremath{\aunit{Me\kern -0.1em V\!/}c}\xspace}
\newcommand{\gevc}{\ensuremath{\aunit{Ge\kern -0.1em V\!/}c}\xspace}
\newcommand{\mevcc}{\ensuremath{\aunit{Me\kern -0.1em V\!/}c^2}\xspace}
\newcommand{\gevcc}{\ensuremath{\aunit{Ge\kern -0.1em V\!/}c^2}\xspace}
\def\fb   {\ensuremath{\aunit{fb}}\xspace}
\def\invfb   {\ensuremath{\fb^{-1}}\xspace}

\def\gsim{{~\raise.15em\hbox{$>$}\kern-.85em
          \lower.35em\hbox{$\sim$}~}\xspace}
\def\lsim{{~\raise.15em\hbox{$<$}\kern-.85em
          \lower.35em\hbox{$\sim$}~}\xspace}

\newcommand{\Real}{\ensuremath{\mathcal{R}e}\xspace}
\newcommand{\Imag}{\ensuremath{\mathcal{I}m}\xspace}

\def\mrad{\aunit{mrad}\xspace}

\def\tell1  {TELL1\xspace}
\def\ukl1   {UKL1\xspace}

\newcommand{\ie}{\mbox{\itshape i.e.}\xspace}

\hypersetup{
  colorlinks   = true, %Colours links instead of ugly boxes
  urlcolor     = blue, %Colour for external hyperlinks
  linkcolor    = blue, %Colour of internal links
  citecolor    = red   %Colour of citations
}

\ifEnableSectionTOCLinks
    \usepackage[explicit]{titlesec} % to change headings
    
    \let\oldcontentsline\contentsline
    \renewcommand

    \titleformat{\section}{\normalfont\Large\bf}{\hyperlink{tocsection.\thesection}{{\thesection} \parbox[t]{\dimexpr\textwidth-1pc}{#1}}}{1pc}{}

    \titleformat{\subsection}{\normalfont\bf}{\hyperlink{tocsubsection.\thesubsection}{{\thesubsection} \parbox[t]{\dimexpr\textwidth-1pc}{#1}}}{1pc}{}

    \titleformat{name=\section,numberless}[display]{}{}{0pt}{\normalfont\Huge\bfseries #1}
\fi

\usepackage{cite} % Allows for ranges in citations
\usepackage{mciteplus}
\def\BdsToKSKPi {\ensuremath{\Bds \to \KS K^\pm \pi^\mp}\xspace}
\def\BdToKSKPi {\ensuremath{\Bd \to \KS K^\pm \pi^\mp}\xspace}
\def\BsToKSKPi {\ensuremath{\Bs \to \KS K^\pm \pi^\mp}\xspace}

\def\forfbar {\kern \thebaroffset\optbar{\kern -\thebaroffset f}{}\xspace}

\usepackage{longtable} % only for template; not usually to be used in PAPERs

\begin{document}

%%%%%%%%%%%%%%%%%%%%%%%%%
%%%%% Title     %%%%%%%%%
%%%%%%%%%%%%%%%%%%%%%%%%%
\renewcommand{\thefootnote}{\fnsymbol{footnote}}
\setcounter{footnote}{1}

% %%%%%%% CHOOSE TITLE PAGE--------
%\onecolumn
%\input{title-LHCb-INT}
%\input{title-LHCb-ANA}
%\input{title-LHCb-CONF}
%\input{title-LHCb-FIGURE}
% ===============================================================================
% Purpose: LHCb-PAPER journal paper title page template
% Author: 
% Created on: 2010-09-25
% ===============================================================================

%%%%%%%%%%%%%%%%%%%%%%%%%
%%%%%  TITLE PAGE  %%%%%%
%%%%%%%%%%%%%%%%%%%%%%%%%
\begin{titlepage}
\pagenumbering{roman}

\vspace*{4.0cm}

% Title --------------------------------------------------
{\normalfont\bfseries\boldmath\huge
\begin{center}
% DO NOT EDIT HERE. Instead edit macro in main.tex to keep metadata correct
  \papertitle 
\end{center}
}

\vspace*{2.0cm}

% Authors -------------------------------------------------
\begin{center}
%In the footnote, replace 'paper' by 'Letter' in case of submission to PRL or PLB 
% Edit macro in main.tex to keep metadata correct
\paperauthors

\textit{University of Warwick, Coventry, United Kingdom}
\end{center}

\vspace{\fill}

% Abstract -----------------------------------------------
\begin{abstract}
  \noindent
The \BsToKSKPi decays are of interest to test the Standard Model and search for new sources of \CP violation.
A full study of these decays requires a tagged decay-time-dependent Dalitz-plot analysis performed simultaneously in the two final states.
Such an analysis has never previously been performed.
The method to carry out such an analysis, relating the amplitudes for decays to resonances in the two final-states to each other, is set out and its feasibility is demonstrated using pseudoexperiments.
The sensitivity to the weak phase difference, $\phi_s^{\rm eff}$, between \decay{\Bs}{\KS\!\KorKbar\!(892)^{*0}} decays with and without \Bs--\Bsb mixing is studied. 
Good precision on $\phi_s^{\rm eff}$ is found to be achievable with a dataset corresponding to LHCb Runs~1--3, with further improvement expected with datasets to be collected in future. 
The method is implemented in the \texttt{Laura++} Dalitz-plot analysis package, and can be applied to other multibody decays with multiple final states.
\end{abstract}

\vspace{\fill}

\end{titlepage}

%%%%%%%%%%%%%%%%%%%%%%%%%%%%%%%%
%%%%%  EOD OF TITLE PAGE  %%%%%%
%%%%%%%%%%%%%%%%%%%%%%%%%%%%%%%%

%  empty page follows the title page ----
\newpage
\setcounter{page}{2}
\mbox{~}

%\twocolumn
% %%%%%%%%%%%%% ---------

\renewcommand{\thefootnote}{\arabic{footnote}}
\setcounter{footnote}{0}

%%%%%%%%%%%%%%%%%%%%%%%%%%%%%%%%
%%%%%  Table of Content   %%%%%%
%%%%%%%%%%%%%%%%%%%%%%%%%%%%%%%%
%%%% Uncomment if desired
%\tableofcontents

\cleardoublepage

%%%%%%%%%%%%%%%%%%%%%%%%%
%%%%% Main text %%%%%%%%%
%%%%%%%%%%%%%%%%%%%%%%%%%

\pagestyle{plain} % restore page numbers for the main text
\setcounter{page}{1}
\pagenumbering{arabic}

%% Uncomment during review phase. 
%% Comment before a final submission.
% \linenumbers

%% This is the main body
%% It is useful to have a single file so comments are not missed in overleaf.
\section{Introduction}
\label{sec:Introduction}

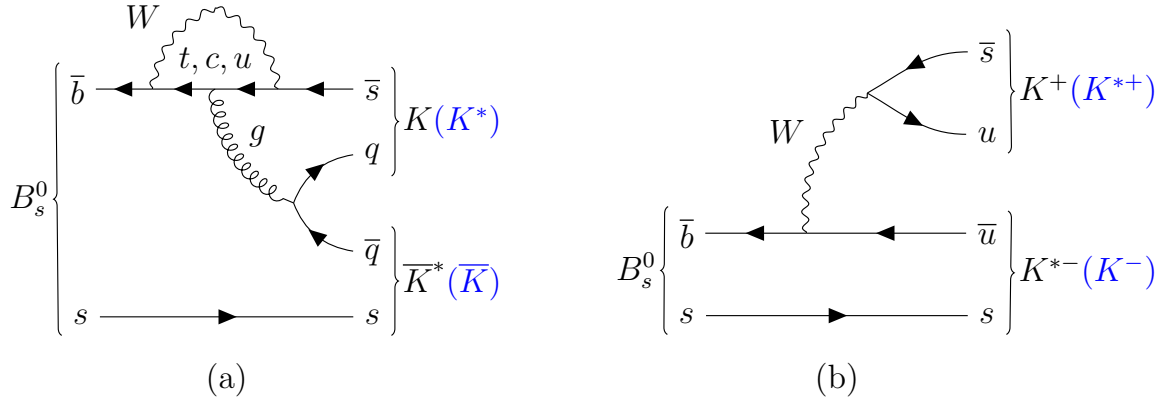
\begin{figure}[bp]
\centering
\resizebox{0.495\textwidth}{!}{
\begin{tikzpicture}
    
\begin{feynman}
\vertex (bbar) {\(\overline b\)};
\vertex[right=1.75cm of bbar] (vtt);
\vertex[right=1.75cm of vtt] (sbar) {\(\overline s \)};

\vertex[left=0.8cm of vtt] (vtb);
\vertex[right=0.8cm of vtt] (vts);
\vertex[above=2.5em of vtt] (vw);
\vertex[above=0.1em of vtt] (tcu) {\(t,c,u\)};

\vertex[below=2.0em of sbar] (u) {\(q\)};
\vertex[below=3.0em of u] (ubar) {\(\overline q\)};
\vertex[below=2.0em of ubar] (sf) {\(s \)};

\vertex[left=3.7cm of sf] (si) {\(s\)};

% \vertex[above=3.0em of ubar] (u) {\(u\)};
% \vertex[above=2.5em of u] (sbar) {\(\overline s\)};

\vertex at ($(ubar)!0.5!(u) - (1.0cm, 0)$) (vuu);

\vertex at ($(sf)!0.5!(si) - (0, 0.8cm)$) (label) {(a)};

\diagram* {
(ubar) -- [fermion, bend left] (vuu) -- [fermion, bend left] (u),
(si) -- [fermion] (sf),
(sbar) -- [fermion] (vts) -- [fermion] (vtt) -- [fermion] (vtb) -- [fermion] (bbar),
(vtt) -- [gluon, bend right, edge label=\(g\)] (vuu),
(vtb) -- [boson, bend left, edge label=\(W\)] (vw) -- [boson, bend left] (vts),
};

\draw [decoration={brace}, decorate] (si.south west) -- (bbar.north west)  node [pos=0.5, left] {\(B_s^{0}\)};

\draw [decoration={brace}, decorate] (ubar.north east) -- (sf.south east) 
node [pos=0.5, right] {\(\overline K^{*} \textcolor{blue}{(\overline K)} \phantom{{}^{*++}} \)};

\draw [decoration={brace}, decorate] (sbar.north east) -- (u.south east) node [pos=0.5, right] {\(K \textcolor{blue}{(K^*)} \phantom{{}^{*++}} \)};

\end{feynman}

\end{tikzpicture}
}
\resizebox{0.495\textwidth}{!}{
\begin{tikzpicture}
    
\begin{feynman}
\vertex (bbar) {\(\overline b\)};
\vertex[right=1.5cm of bbar] (vub);
\vertex[right=2.0cm of vub] (ubar) {\(\overline u\)};

\vertex[below=2.5em of bbar] (si) {\(s\)};
\vertex[below=2.5em of ubar] (sf) {\(s\)};

\vertex[above=3.0em of ubar] (u) {\(u\)};
\vertex[above=2.5em of u] (sbar) {\(\overline s\)};

\vertex at ($(sbar)!0.5!(u) - (1.5cm, 0)$) (vus);

\vertex at ($(sf)!0.5!(si) - (0, 0.8cm)$) (label) {(b)};

\diagram* {
(ubar) -- [fermion] (vub) -- [fermion] (bbar),
(si) -- [fermion] (sf),
(sbar) -- [fermion, out=180, in=30] (vus) -- [fermion, out=-30, in=180] (u),
(vub) -- [boson, bend left, edge label=\(W\)] (vus),
};

\draw [decoration={brace}, decorate] (si.south west) -- (bbar.north west) node [pos=0.5, left] {\(B_s^{0}\)};

\draw [decoration={brace}, decorate] (ubar.north east) -- (sf.south east) node [pos=0.5, right] {\(K^{*-} \textcolor{blue}{(K^{-})}\)};

\draw [decoration={brace}, decorate] (sbar.north east) -- (u.south east) node [pos=0.5, right] {\(K^{+} \textcolor{blue}{(K^{*+})}\)};

\end{feynman}

\end{tikzpicture}
}

\caption{\small The SM (a)~loop and (b)~tree diagrams contributing to the $\Bs\to\KS\Kpm\pimp$ decays.  The quarks $q$ can either be $u$ or $d$. The $\KS\Kmp\pipm$ final states can be accessed from $\Kstarz\to \Kp\pim$ and $\Kstarp\to\Kz\pip$, followed by $\Kz\to\KS$.
Swapping all $s$ and $d$ quarks in (a) results in the corresponding diagrams for the $\Bd\to\KS\Kpm\pimp$ decays.}
\label{fig:feynman}
\end{figure}

Studies of \CP violation play a key role in testing the Standard Model~(SM) of particle physics and probing possible new physics~(NP) beyond it. 
Charmless $B$-meson decays provide plenty of opportunities in this endeavour.
Many decays in this category involve, at the quark level, flavour-changing neutral-current~(FCNC) $b \to s/d$ transitions, which proceed only via suppressed loop~(penguin) diagrams in the SM, as shown in Fig.~\ref{fig:feynman}(a). As a result, these decays are sensitive to possible NP contributions.

The \BsToKSKPi decays~\footnote{Charge-conjugate processes are implicitly included throughout unless stated otherwise.} are among the channels of particular interest.
When an intermediate $\KorKbar\!^{*0}$ resonance is involved, the decay is dominated in the SM by a single $t$-loop penguin diagram, as illustrated in Fig.~\ref{fig:feynman}(a).
Therefore there is neither \CP violation in the decay nor, since the weak phase difference between decays with and without \Bs--\Bsb oscillation vanishes, any mixing-induced \CP violation. 
Contributions from higher-order $c$- and $u$-loop diagrams may modify this expectation.
Their impact can, however, be constrained by relating the observables to those in the corresponding $\Bd$ decays using $U$-spin symmetry, as discussed in the literature for \decay{\BdorBs}{\KS\KS} and \Kstarz\Kstarzb decays~\cite{Descotes-Genon:2006spp,Ciuchini:2007hx,Bhattacharya:2012hh}.
Since the latest experiment results for both \decay{\BdorBs}{\KS\KS}~\cite{BaBar:2006enb,Belle:2007cga,Belle:2012dmz,Belle:2015gho,LHCb-PAPER-2019-030} and \decay{\BdorBs}{\Kstarz\Kstarzb}~\cite{LHCb-PAPER-2017-048,LHCb-PAPER-2025-046} decays exhibit some disagreements with SM predictions~\cite{Alguero:2020xca,Amhis:2022hpm,Biswas:2023pyw,Aleksan:2023rkc}, investigation of the corresponding decays to $\KS\Kstarz$ and $\KS\Kstarzb$ are well motivated.

The \BsToKSKPi decays also include contributions from \Kstarpm resonances.
In this case, an additional tree-level diagram can contribute , as shown in Fig.~\ref{fig:feynman}(b). 
This makes \CP violation in decay possible, and provides sensitivity to the CKM angle $\gamma$ if the contributions from the loop and tree diagrams can be disentangled.

Access to the full set of \CP-violating observables in \BsToKSKPi decays requires a flavour-tagged decay-time-dependent Dalitz-plot analysis.
% To properly disentangle the contributions of intermediate resonances, a Dalitz-plot analysis is also needed. The interference between resonances leads to varying strong-phase differences across the Dalitz plot. This can enhance the sensitivity to \CP-violating observables compared to the quasi-two-body approach that treats the \Kstar resonance, which has a non-negligible natural width, as a final-state particle. 
This approach has already been employed in data analysis of the three-body charmless decays $\Bd\to\KS\pip\pim$~\cite{PhysRevD.80.112001,PhysRevD.79.072004}, $\Bd\to\KS\Kp\Km$~\cite{PhysRevD.82.073011,PhysRevD.85.112010} and $\Bd\to\pip\pim\piz$~\cite{BaBar:2013uwm,Belle:2007jkw,Belle:2007krm} by the BaBar and Belle collaborations, as well as the four-body $\Bs\to (\Kp\pim)(\Km\pip)$~\cite{LHCb-PAPER-2017-048} and $\Bs\to (\Kp\Km)(\Kp\Km)$~\cite{LHCb-PAPER-2023-001} decays by LHCb. 
A sensitivity study based on pseudoexperiments has also been conducted for the $\Bs\to\KS\pip\pim$ channel~\cite{Gershon:2014yma}.
%and $\Bd\to\Dz\pip\pim$ channels~\cite{Latham_2009}. 
All of these channels involve a single self-conjugate final state, whereas the analysis of the \BsToKSKPi decays requires a simultaneous analysis of two final states. 
Although more complex than the single-final-state case, constraints from relations between the two final states provide additional sensitivity.
Similar methods exploiting relations between two or more Dalitz plots have been exploited in some analyses without decay-time-dependence~\cite{ LHCb-PAPER-2015-059,LHCb-PAPER-2022-026,LHCb-PAPER-2022-027,LHCb-PAPER-2023-047}, but no simultaneous decay-time-dependent Dalitz-plot analysis of multiple final states has previously been carried out.

The \BdsToKSKPi decays have been observed by LHCb, and their branching fractions have been measured~\cite{LHCb-PAPER-2013-042,LHCb-PAPER-2014-043,LHCb-PAPER-2015-018,LHCb-PAPER-2024-029}. An untagged time-integrated Dalitz-plot analysis has been performed for the \Bs channel~\cite{LHCb-PAPER-2018-045}. Due to the limited data statistics collected during the LHC operation periods Run\,1~(2011--2012) and Run\,2~(2015--2018), no flavour-tagged analysis has yet been carried out. The large dataset being collected by LHCb in Run\,3~(2022--2026) is expected to enable a flavour-tagged analysis of the \BsToKSKPi decays, with more precise measurements foreseen with larger data samples in future.
No other current or near-term experiment has the capability to carry out decay-time-dependent Dalitz-plot analysis of \BdsToKSKPi decays.
Further in the future, experiments at the proposed Future Circular Collider could have interesting sensitivity.

Given the novelty and complexity of a simultaneous flavour-tagged decay-time-dependent Dalitz-plot analysis in two final states, a sensitivity study based on pseudoexperiments is performed in this paper. The implementation is based on the \texttt{Laura++} package~\cite{Back:2017zqt}. The achievable sensitivities to the \CP-violating observables for LHCb Run\,3 data and future operation periods are evaluated, providing guidance for future data analyses. This study also establishes a baseline analysis framework for this class of measurements.

The remainder of the paper is organised as follows.
In Sec.~\ref{sec:Formalism} an overview of the formalism is given, with details of the implementation used in the pseudoexperiments presented in Sec.~\ref{sec:Pseudoexperiments}.
The results are presented and discussed in Sec.~\ref{sec:Results}.
A brief summary concludes the paper in Sec.~\ref{sec:Summary}.

\section{Formalism}
\label{sec:Formalism}
Following the convention of Ref.~\cite{Anikeev:2001rk}, the differential decay widths for mesons initially produced as \Bsb and \Bs, decaying to the final states $f\equiv\KS\Kp\pim$ and $\bar{f}\equiv\KS\Km\pip$, can be expressed as functions of the decay time $t$ and the phase-space variables $\Phi_3$~\footnote{\label{fn:DP-def}Since the two final-states contain different particles, strictly speaking their phase spaces are different; nonetheless the same symbol $\Phi_3$ is used for both.  Specifically, $\Phi_3\equiv(m^2_{\KS\pim},\,m^2_{\Kp\pim})$ for the $\KS\Kp\pim$ final state, and $\Phi_3 \equiv (m^2_{\KS\pip},\,m^2_{\Km\pip})$ for the $\KS\Km\pip$ final state.}
as

\begin{equation}
  \label{eq:cp_uta:td_cp_bsb_asp1}
  \begin{array}{lcr}
    \multicolumn{2}{l}{
      \frac{d^2}{dtd\Phi_3} \Gamma_{\Bsb \to f} (t) =      
      \frac{{\cal N} \, e^{-t / \tau(\Bs)}}{2\tau(\Bs)}
      \Big[ 
      \left( \left| {\cal A}_f \right|^2 + \left| \bar{\cal A}_f \right|^2 \right) \cosh\left(\frac{\DGs t}{2}\right) +
      2\,\Imag\left( \frac{q}{p} {\cal A}_f^* \bar{\cal A}_f \right) \sin(\dms t) - {}
    } & \\
    &
    \multicolumn{2}{r}{
      \left( \left| {\cal A}_f \right|^2 - \left| \bar{\cal A}_f \right|^2 \right) \cos(\dms t) -
      2\,\Real\left( \frac{q}{p} {\cal A}_f^* \bar{\cal A}_f \right) \sinh\left(\frac{\DGs t}{2}\right)
      \Big]\,,
    } \\
  \end{array}
\end{equation}
\begin{equation}
  \label{eq:cp_uta:td_cp_bs_asp1}
  \begin{array}{lcr}
    \multicolumn{2}{l}{
      \frac{d^2}{dtd\Phi_3} \Gamma_{\Bs \to f} (t) =  
      \frac{{\cal N} \, e^{-t / \tau(\Bs)}}{2\tau(\Bs)}
      \Big[ 
      \left( \left| {\cal A}_f \right|^2 + \left| \bar{\cal A}_f \right|^2 \right) \cosh \left(\frac{\DGs t}{2}\right) -
      2\,\Imag\left( \frac{q}{p} {\cal A}_f^* \bar{\cal A}_f \right) \sin(\dms t) + {}
    } & \\
    & 
    \multicolumn{2}{r}{
      \left( \left| {\cal A}_f \right|^2 - \left| \bar{\cal A}_f \right|^2 \right) \cos(\dms t) -
      2\,\Real\left( \frac{q}{p} {\cal A}_f^* \bar{\cal A}_f \right) \sinh\left(\frac{\DGs t}{2}\right)
      \Big]\,,  
    } \\
  \end{array}
\end{equation}
\begin{equation}
  \label{eq:cp_uta:td_cp_bsb_asp2}
  \begin{array}{lcr}
    \multicolumn{2}{l}{
      \frac{d^2}{dt d\Phi_3} \Gamma_{\Bsb \to \bar{f}} (t) =      
      \frac{{\cal N} \, e^{-t / \tau(\Bs)}}{2\tau(\Bs)}
      \Big[ 
      \left( \left| {\cal A}_{\bar{f}} \right|^2 + \left| \bar{\cal A}_{\bar{f}} \right|^2 \right) \cosh\left(\frac{\DGs t}{2}\right) +
      2\,\Imag\left( \frac{q}{p} {\cal A}_{\bar{f}}^* \bar{\cal A}_{\bar{f}} \right) \sin(\dms t) - {}
    } & \\
    &
    \multicolumn{2}{r}{
      \left( \left| {\cal A}_{\bar{f}} \right|^2 - \left| \bar{\cal A}_{\bar{f}} \right|^2 \right) \cos(\dms t) -
      2\,\Real\left( \frac{q}{p} {\cal A}_{\bar{f}}^* \bar{\cal A}_{\bar{f}} \right) \sinh\left(\frac{\DGs t}{2}\right)
      \Big]\,,
    } \\
  \end{array}
\end{equation}
\begin{equation}
  \label{eq:cp_uta:td_cp_bs_asp2}
  \begin{array}{lcr}
    \multicolumn{2}{l}{
      \frac{d^2}{dt d\Phi_3} \Gamma_{\Bs \to \bar{f}} (t) =  
      \frac{{\cal N} \, e^{-t / \tau(\Bs)}}{2\tau(\Bs)}
      \Big[ 
      \left( \left| {\cal A}_{\bar{f}} \right|^2 + \left| \bar{\cal A}_{\bar{f}} \right|^2 \right) \cosh \left(\frac{\DGs t}{2}\right) -
      2\,\Imag\left( \frac{q}{p} {\cal A}_{\bar{f}}^* \bar{\cal A}_{\bar{f}} \right) \sin(\dms t) + {}
    } & \\
    & 
    \multicolumn{2}{r}{
      \left( \left| {\cal A}_{\bar{f}} \right|^2 - \left| \bar{\cal A}_{\bar{f}} \right|^2 \right) \cos(\dms t) -
      2\,\Real\left( \frac{q}{p} {\cal A}_{\bar{f}}^* \bar{\cal A}_{\bar{f}} \right) \sinh\left(\frac{\DGs t}{2}\right)
      \Big]\,,  
    } \\
  \end{array}
\end{equation}
where $\bar{\cal A}_f$ and ${\cal A}_f$ denote the decay amplitudes for \Bsb and \Bs decays to the final state $f$, respectively, while $\bar{\cal A}_{\bar{f}}$ and ${\cal A}_{\bar{f}}$ are defined analogously for $\bar{f}$. All amplitudes depend on the Dalitz-plot variables $\Phi_3$.
The mass and width differences between the light (L) and heavy (H) \Bs mass eigenstates are defined as $\dms = m_{\rm H} - m_{\rm L}$ and $\DGs = \Gamma_{\rm L} - \Gamma_{\rm H}$, and the \Bs lifetime is $\tau(\Bs) = \left(\frac{\Gamma_{\rm L} + \Gamma_{\rm H}}{2}\right)^{-1}$. The mass eigenstates are given by $| B^0_{s\,{\rm L}} \rangle = p | \Bs \rangle + q | \Bsb \rangle$ and $| B^0_{s\,{\rm H}} \rangle = p | \Bs \rangle - q | \Bsb \rangle$, with $\left| p \right|^2 + \left| q \right|^2 = 1$. The \CP\ phase convention ($\CP| \Bs \rangle = | \Bsb \rangle$) is chosen such that the lighter eigenstate is \CP-even in the limit $\left| q/p \right| = 1$, which is a good approximation experimentally~\cite{LHCb-PAPER-2016-013}.
The normalisation factor $\mathcal{N}$ is defined by requiring the total probability density to be unity. It is given by the inverse of the quantity obtained after integrating over the decay time and Dalitz-plot variables and summing over the initial and final states, as
\begin{equation}
{\cal N}  = \frac{1}{
    \int \left( \left| {\cal A}_f \right|^2 + \left| \bar{\cal A}_f \right|^2 \right) \left( \frac{1+yA^{\DGs}_f}{1-y^2} \right){\rm d}\Phi_3
    + 
    \int \left( \left| {\cal A}_{\bar{f}} \right|^2 + \left| \bar{\cal A}_{\bar{f}} \right|^2 \right) \left( \frac{1+yA^{\DGs}_{\bar{f}}}{1-y^2} \right){\rm d}\Phi_3
    }\,.
\label{eq:norm2}
\end{equation}
where $y=\Delta\Gamma/2\Gamma\approx 0.065\pm0.005$~\cite{PDG2024}, $A_f^{\Delta\Gamma_s} \equiv -2\,\Real\left(\frac{q}{p}\mathcal{A}^*_f \bar{\mathcal{A}}_f\right)/\left(|\mathcal{A}_f|^2+|\bar{\mathcal{A}}_f|^2\right)$ and $A_{\bar{f}}^{\Delta\Gamma_s}$ is defined similarly with all $f \to \bar{f}$.

\CP violation can manifest in these decay-time and Dalitz-plot-dependent decay rates in different ways.
\CP violation in decay appears as a difference between the magnitudes of an amplitude and its \CP-conjugate, \ie\ between $\mathcal{A}_f$ and $\bar{\mathcal{A}}_{\bar{f}}$, or between $\mathcal{A}_{\bar{f}}$ and $\bar{\mathcal{A}}_{f}$, at specific points in phase space.
Such \CP-violation effects cause the cosine and hyperbolic cosine terms of Eqs.~\eqref{eq:cp_uta:td_cp_bsb_asp1} and~\eqref{eq:cp_uta:td_cp_bs_asp1} to differ from those of Eqs.~\eqref{eq:cp_uta:td_cp_bsb_asp2} and~\eqref{eq:cp_uta:td_cp_bs_asp2}.
\CP violation in the interference between mixing and decay is reflected in the sine terms, which appear with opposite signs between Eqs.~\eqref{eq:cp_uta:td_cp_bsb_asp1} and~\eqref{eq:cp_uta:td_cp_bs_asp1}, and likewise between Eqs.~\eqref{eq:cp_uta:td_cp_bsb_asp2} and~\eqref{eq:cp_uta:td_cp_bs_asp2}, and hence cause decay-time-dependent asymmetries.
The total phase difference~($\phi^{\rm eff}_s$) that appears in these coefficients is the sum of contributions from \Bs--\Bsb mixing, $\arg\left(\frac{q}{p}\right)$, and decay, $\arg\left(2\,\mathcal{A}^*_f \bar{\mathcal{A}}_f\right)/\left(|\mathcal{A}_f|^2+|\bar{\mathcal{A}}_f|^2\right)$ or the corresponding expression with $f \to \bar{f}$.
Since the phase difference related to the decay amplitudes can contain both strong and weak phase components, the sine coefficients being nonzero is not in itself an unambiguous sign of \CP violation.
Instead, \CP violation occurs if the sum of the coefficients from the two pairs of equations, at a specific point in phase-space, is nonzero.  
The hyperbolic sine terms are sensitive to the same phase differences, but these appear with the same sign for all decays and hence cause changes to the effective lifetime rather than asymmetries.  

The decay amplitudes are usually constructed within the isobar formalism, in which the three-body decay amplitude is expressed as a coherent sum of contributions from amplitudes associated to sequential two-body decays, 
\begin{equation}
    \label{eq:isobar}
    {\cal A}_f = \sum_i c_i F_i(\Phi_3)\,.
\end{equation}
Here the sum runs over intermediate resonant and nonresonant contributions, with complex coefficients $c_i$ and dynamics including mass- and angular-dependent terms described by $F_i(\Phi_3)$.
The total amplitude for decay to the charge-conjugate final state, ${\cal A}_{\bar{f}}$, can be constructed similarly with a sum over the corresponding charge-conjugate intermediate states.
In the limit of \CP symmetry, ${\cal A}_f = \bar{\cal A}_{\bar{f}}$ and ${\cal A}_{\bar{f}} = \bar{\cal A}_f$, so relations can be imposed between the complex coefficients associated with the different amplitudes, as shown in the next section.

\section{Pseudoexperiment setup}
\label{sec:Pseudoexperiments}

In order to generate and study pseudoexperiments, a choice of model --- amounting to a choice of the resonant and nonresonant terms to include in the sum in Eq.\eqref{eq:isobar} and corresponding expressions for the other amplitudes --- needs to be made.
A simple model is chosen in which the only resonances included are the neutral and charged $\Kstar(892)$ states.~\footnote{The symbol $\Kstar$ denotes the $\Kstar(892)$ resonance hereafter, unless stated otherwise.} 
The four amplitudes can then be written explicitly as
\begin{align}
{\cal A}_f &= c_{\Kstarz} F_{\Kstarz}(\Phi_3) + c_{\Kstarm} F_{\Kstarm}(\Phi_3)\,, \\
\bar{\cal A}_f &= \bar{c}_{\Kstarz} F_{\Kstarz}(\Phi_3) + \bar{c}_{\Kstarm} F_{\Kstarm}(\Phi_3)\,, \\
{\cal A}_{\bar{f}} &= c_{\Kstarzb} F_{\Kstarzb}(\Phi_3) + c_{\Kstarp} F_{\Kstarp}(\Phi_3)\,, \\
\bar{\cal A}_{\bar{f}} &= \bar{c}_{\Kstarzb} F_{\Kstarzb}(\Phi_3) + \bar{c}_{\Kstarp} F_{\Kstarp}(\Phi_3)\,.
\end{align}
It is convenient to decompose the complex coefficients into \CP-conserving and \CP-violating parts, and thus relating the terms that appear in ${\cal A}_f$ with those in $\bar{\cal A}_{\bar{f}}$ and likewise relating terms that appear in ${\cal A}_{\bar{f}}$ with those in $\bar{\cal A}_f$.
Explicitly,
\begin{equation}
\begin{array}{l @{\quad} l}
c_{\Kstarz} = c_0 + \Delta c_0\,, & \bar{c}_{\Kstarzb} = c_0 - \Delta c_0\,,\\
c_{\Kstarm} = c_- + \Delta c_-\,, & \bar{c}_{\Kstarp} = c_- - \Delta c_-\,,\\
c_{\Kstarzb} = c_{\bar{0}} + \Delta c_{\bar{0}}\,, & \bar{c}_{\Kstarz} = c_{\bar{0}} - \Delta c_{\bar{0}}\,,\\
c_{\Kstarp} = c_{\bar{-}} + \Delta c_{\bar{-}}\,, & \bar{c}_{\Kstarm} = c_{\bar{-}} - \Delta c_{\bar{-}}\,,
\end{array}
\end{equation}
where the notation $c_{\bar{-}}$ is used, rather than $c_+$, for consistency with the $c_{\bar{0}}$ notation. 

The parameters $\Delta c_{0,-,\bar{0},\bar{-}}$ encode \CP violation as differences between \CP-conjugate decay amplitudes, while the $c_{0,-,\bar{0},\bar{-}}$ parameters are \CP conserving.
Each $c$ and $\Delta c$ term is complex and can be expressed in Cartesian form, \ie\ in terms of real and imaginary parts, as 
\begin{equation}
\begin{array}{l @{\quad} l}
c_0 = x_0 + i y_0 \,, & \Delta c_0 = \Delta x_0 + i \Delta y_0 \,,\\
c_- = x_- + i y_- \,, & \Delta c_- = \Delta x_- + i \Delta y_- \,,\\
c_{\bar{0}} = x_{\bar{0}} + i y_{\bar{0}} \,, & \Delta c_{\bar{0}} = \Delta x_{\bar{0}} + i \Delta y_{\bar{0}} \,,\\
c_{\bar{-}} = x_{\bar{-}} + i y_{\bar{-}} \,, & \Delta c_{\bar{-}} = \Delta x_{\bar{-}} + i \Delta y_{\bar{-}} \,.
\end{array}
\end{equation}

It is worth emphasising that $c_0$ and $c_{\bar{0}}$ and likewise $c_-$ and $c_{\bar{-}}$, are independent parameters, which are not related by \CP.  
That can be seen by noting that, for example, $c_{\Kstarz}$ and $c_{\Kstarzb}$ are related to the rate at which the $\bar{q}$ quark hadronises together with the spectator $s$ quark in Fig.~\ref{fig:feynman}(a) to form a \KS or a $\Kstarzb$ meson, respectively.

This parameterisation is implemented in the \texttt{Laura++} package~\cite{Back:2017zqt}, which is used to generate and fit the pseudoexperiments. 
The mass-dependent term in $F_{\Kstar}(\Phi_3)$ is described by a relativistic Breit--Wigner~(BW) function with appropriate barrier factors, with the \Kstar mass and width fixed to their known values~\cite{PDG2024}.
The angular-dependent term is described using the Zemach formalism~\cite{Zemach:1963bc,Zemach:1965ycj}.

In the baseline model, all $c_i$ are set to unity and all $\Delta c_i$ are set to zero.
This corresponds to a choice of phase convention in which $\phi_s^{\rm eff}$ comes only from $\arg\left(\frac{q}{p}\right)$; this phase is set to $-2\beta_s = -0.036$~\cite{PDG2024}, where $\beta_s = \arg\left(-\frac{V_{ts}^{}V_{tb}^*}{V_{cs}^{}V_{cb}^*}\right)$.
A set of variations around this baseline model is investigated.
In all variations, the $\Bs\to\Kstarm\Kp$ amplitude, contributing to $\mathcal{A}_f$~($\bar{\mathcal{A}}_{\bar{f}}$), is taken as the reference, with $c_-$ unvaried. The explicit parameter values for each scenario are listed in Table~\ref{tab:configs}, and correspond to:

\begin{enumerate}[start=1]
\item Magnitude variations:
\begin{enumerate}
    \item Varying $|c_i|$ for the $\Bs \to \KS\Kstarzb$ and $\Bs \to \Kstarp \Km$ amplitudes, both contributing to $\mathcal{A}_{\bar{f}}$~($\bar{\mathcal{A}}_f$);
    \item Varying $|c_i|$ for the $\Bs\to \KS\Kstarz$ and $\Bs\to\KS\Kstarzb$ amplitudes, contributing to $\mathcal{A}_f$~($\bar{\mathcal{A}}_{\bar{f}}$) and $\mathcal{A}_{\bar{f}}$~($\bar{\mathcal{A}}_f$), respectively.
\end{enumerate}

\item Phase variations:
\begin{enumerate}
    \item Varying $\arg(c_i)$ for the $\Bs\to\Kstarp\Km$ amplitude, contributing to $\mathcal{A}_{\bar{f}}$~($\bar{\mathcal{A}}_f$);
    \item Varying $\arg(c_i)$ for the $\Bs\to\KS\Kstarz$ amplitude, contributing to $\mathcal{A}_f$~($\bar{\mathcal{A}}_{\bar{f}}$).
\end{enumerate}

\item Introducing \CP violation in $\Kstarpm$ amplitudes, expected from the interference between SM loop and tree diagrams as described in Sec.~\ref{sec:Introduction}:
\begin{enumerate}
    \item Varying $|\Delta c_i|$ for the $\Bs\to\Kstarm\Kp$ amplitude, contributing to $\mathcal{A}_f$~($\bar{\mathcal{A}}_{\bar{f}}$);
    \item Varying $|\Delta c_i|$ for the $\Bs\to\Kstarp\Km$ amplitude, contributing to $\mathcal{A}_{\bar{f}}$~($\bar{\mathcal{A}}_f$).
\end{enumerate}

\item Introducing \CP violation in $\Kstarz$ and $\Kstarzb$ amplitudes, potentially arising from NP as discussed in Sec.~\ref{sec:Introduction}:
\begin{enumerate}
    \item Varying $|\Delta c_i|$ for the $\Bs\to\KS\Kstarz$ amplitude, contributing to $\mathcal{A}_f$~($\bar{\mathcal{A}}_{\bar{f}}$);
    \item Varying $|\Delta c_i|$ for the $\Bs\to\KS\Kstarzb$ amplitude, contributing to $\mathcal{A}_{\bar{f}}$~($\bar{\mathcal{A}}_f$).
\end{enumerate}

\item Variations in $\arg\left(\frac{q}{p}\right)$, effectively equivalent to a common global shift in all $\arg(\Delta c_i)$,~\footnote{In principle, independent variations of $\arg(\Delta c_i)$ could also be investigated, but this would introduce additional complexity and is therefore not pursued.  Similar sensitivity to $\phi_s^{\rm eff}$ is expected in such cases.} since the total weak-phase difference is given by the sum of $\arg\left(\frac{q}{p}\right)$ and the weak phase difference from decay, as discussed in Sec.~\ref{sec:Formalism}. 
%This scenario complements Scenarios~3 and~4. 

\end{enumerate}

\begin{table}[htbp]
\centering
\caption{Model configurations used in pseudoexperiment generation. Parameters not explicitly specified are set to their baseline values.}
\label{tab:configs}
% \begin{tabular}{c p{5cm} p{8cm}}
\begin{tabular}{c @{\hspace{0.5cm}} l @{\hspace{0.7cm}} l}
\hline
No. & Scenario & Configurations \\ 
\hline

0 & Baseline 
& $c_- = c_0 = c_{\bar{0}} = c_{\bar{-}} = 1$ \\
& & $\Delta c_- = \Delta c_0 = \Delta c_{\bar{0}} = \Delta c_{\bar{-}} = 0$ \\
& & $\phi_s^{\rm eff} = -2\beta_s$ \\

\hline

1 & Magnitude variation 
& (a) $c_{\bar{0}} = c_{\bar{-}} = 0.7, 0.8, 0.9$ \\
& & (b) $c_0 = c_{\bar{0}} = 1.1, 1.2, 1.3$ \\

\hline

2 & Phase variation 
& (a) $c_{\bar{-}} = e^{i(\pm \pi/4, \pm \pi/2, \pm 3\pi/4, \pm \pi)}$ \\
& & (b) $c_0 = e^{i(\pm \pi/4, \pm \pi/2, \pm 3\pi/4, \pm \pi)}$ \\

\hline

3 & CPV in $\Kstarpm$ amplitudes
& (a) $\Delta c_- = \pm 0.1, \pm 0.2, \pm 0.3$ \\
& & (b) $\Delta c_{\bar{-}} = \pm 0.1, \pm 0.2, \pm 0.3$ \\

\hline

4 & CPV in $\Kstarz$ amplitudes
& (a) $\Delta c_0 = \pm 0.1, \pm 0.2, \pm 0.3$ \\
& & (b) $\Delta c_{\bar{0}} = \pm 0.1, \pm 0.2, \pm 0.3$ \\

\hline

5 & $\phi_s^{\rm eff}$ variation & $\arg\left(\frac{q}{p}\right) =0,  -10 \beta_s, -40 \beta_s$ \\
& & $c_0 = c_{\bar{0}} = 1.2$ \\ 

\hline
\end{tabular}
\end{table}

Pseudoexperiments are generated using the amplitude models described above, and with statistics corresponding to the combined dataset expected from LHCb Runs~1, 2 and 3 (taking total integrated luminosities of 1, 2, 6 and 23\invfb at $pp$ collision centre-of-mass energies of 7, 8, 13 and 13.6\tev, respectively).
The total yields from Runs~1 and 2 are taken from Ref.~\cite{LHCb-PAPER-2024-029}, while those in Run~3 are scaled accounting for the anticipated efficiency improvement of (at least) a factor of 2 resulting from the upgraded detector and full software trigger~\cite{LHCb-DP-2022-002}.

The total \BsToKSKPi yield is expected to be around $130\,000$, which with a typical effective tagging efficiency of about $5\%$, as observed in similar charmless \Bs decays~\cite{LHCb-PAPER-2017-048,LHCb-PAPER-2018-006,LHCb-PAPER-2023-001}, corresponds to an effective tagged yield of approximately 6500.
The defines the number of \Bs decays generated in each pseudoexperiments. 
Detector effects, including efficiency and resolution and well as imperfect flavour tagging, are not included in the analysis. 
For each amplitude-model configuration, an ensemble of 500 pseudoexperiments is generated.

An unbinned maximum-likelihood method is employed to fit the pseudoexperiments and determine the values and uncertainties of the floating parameters, including all $x_i$, $y_i$, $\Delta x_i$ and $\Delta y_i$ parameters as well as $\phi_s^{\rm eff}$, except for a small subset that is fixed to resolve ambiguities in the fit. 
The fixed parameters and constraints used to resolve ambiguities are:
\begin{enumerate}[start=1]
\item \label{constraint1} $c_-$ is fixed to unity, \ie\ $x_-=1$ and $y_-=0$. This is allowed since an arbitrary complex scale factor can be applied to the differential decay widths given in Eqs.~\eqref{eq:cp_uta:td_cp_bsb_asp1}--\eqref{eq:cp_uta:td_cp_bs_asp2}. The same factor also appears in the normalisation factor of Eq.~\eqref{eq:norm2}, and therefore cancels exactly in the normalised PDF.

\item \label{constraint2} The imaginary part of $\Delta c_-$ is set to zero, \ie\ $\Delta y_-=0$.  With the previous constraint, and assuming $|\Delta x_-|<1$, this corresponds to setting both $\arg\left(c_{\Kstarm}\right)$ and $\arg\left(\bar{c}_{\Kstarp}\right)$ to zero.  This is necessary because an arbitrary global relative phase between $\mathcal{A}_f$ and $\bar{\mathcal{A}}_{\bar{f}}$ can be absorbed into $\phi_s^{\rm eff}$.

\item \label{constraint3} $\Imag\left(\frac{\Delta c_{\bar{-}}}{c_{\bar{-}}}\right)$ is set to zero, \ie\ $x_{\bar{-}}\Delta y_{\bar{-}} - y_{\bar{-}}\Delta x_{\bar{-}}=0$, which is equivalent to setting $\Imag\left(\frac{c_{\Kstarp}}{\bar{c}_{\Kstarm}}\right)$ to zero.  This is necessary in addition to the previous constraint since an independent arbitrary global relative phase between $\mathcal{A}_{\bar{f}}$ and $\bar{\mathcal{A}}_f$ can also be absorbed into $\phi_s^{\rm eff}$. In practice, this requirement is implemented in the fit as a Gaussian constraint on $x_{\bar{-}}\Delta y_{\bar{-}} - y_{\bar{-}}\Delta x_{\bar{-}}$, with mean of zero and an extremely narrow width; this has negligible impact on the parameter uncertainties.

\item \label{constraint4} An additional ambiguity arises from the transformation $\phi_s^{\rm eff} \to \phi_s^{\rm eff} \pm \pi$. This corresponds to extracting a global phase $\mp \pi$ between $\mathcal{A}_f$ and $\bar{\mathcal{A}}_{\bar{f}}$, and simultaneously between $\mathcal{A}_{\bar{f}}$ and $\bar{\mathcal{A}}_f$, and absorbing it into $\phi_s^{\rm eff}$; it occurs because the previous two constraints set the imaginary parts of the amplitudes to be zero, and hence allow phases of $\pm \pi$ as well as zero.  To resolve this ambiguity, only solutions with $\phi_s^{\rm eff}$ within $\pm \pi/3$ of the input value are retained. This selection keeps approximately half of the successful fits for a given model configuration, around 250, which remains sufficient to evaluate the sensitivities of the parameters of interest.
\end{enumerate}
To ensure reliable convergence to the global minimum for each pseudoexperiment, 100 fits are performed from randomly chosen starting points in parameter space. 
To have confidence that the global minimum has been found, at least three successful fits are required to have converged to the best likelihood value obtained in the fits to the ensemble.

\section{Results and discussions}
\label{sec:Results}

\begin{figure}[tbp]
    \centering
    \includegraphics[width=0.49\linewidth]{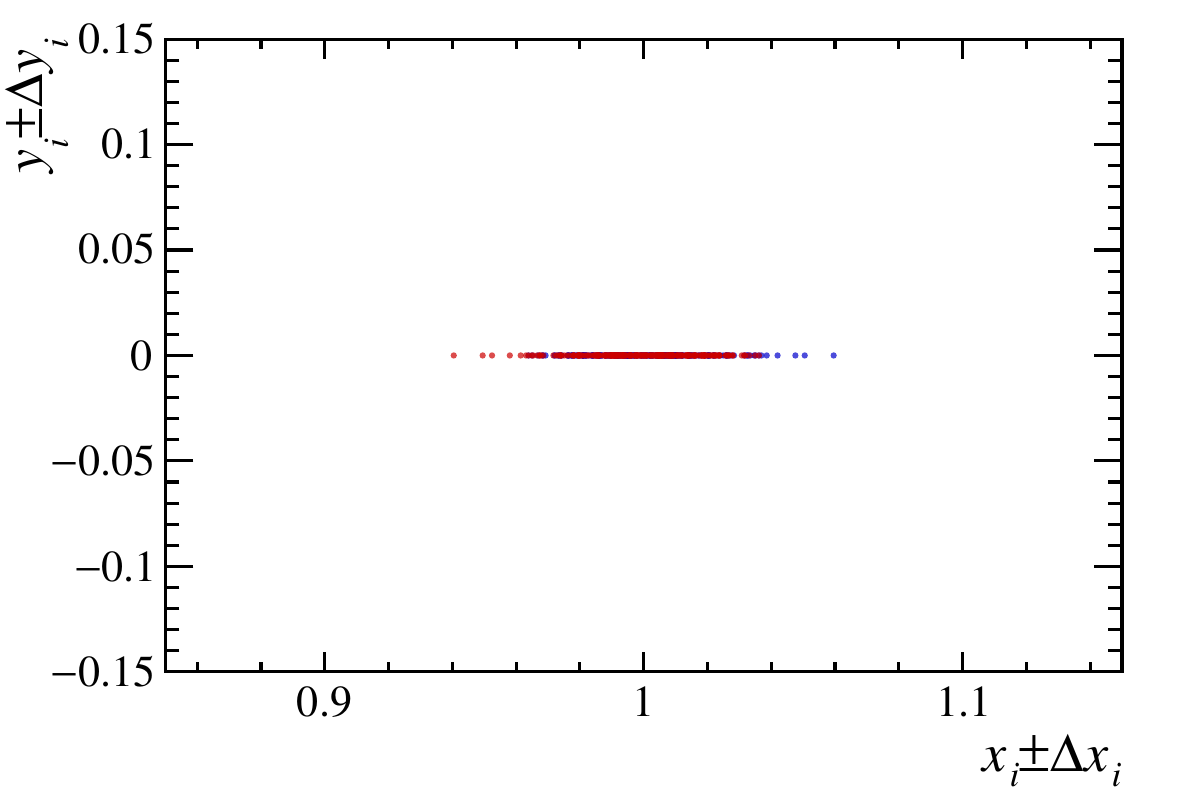}
    \includegraphics[width=0.49\linewidth]{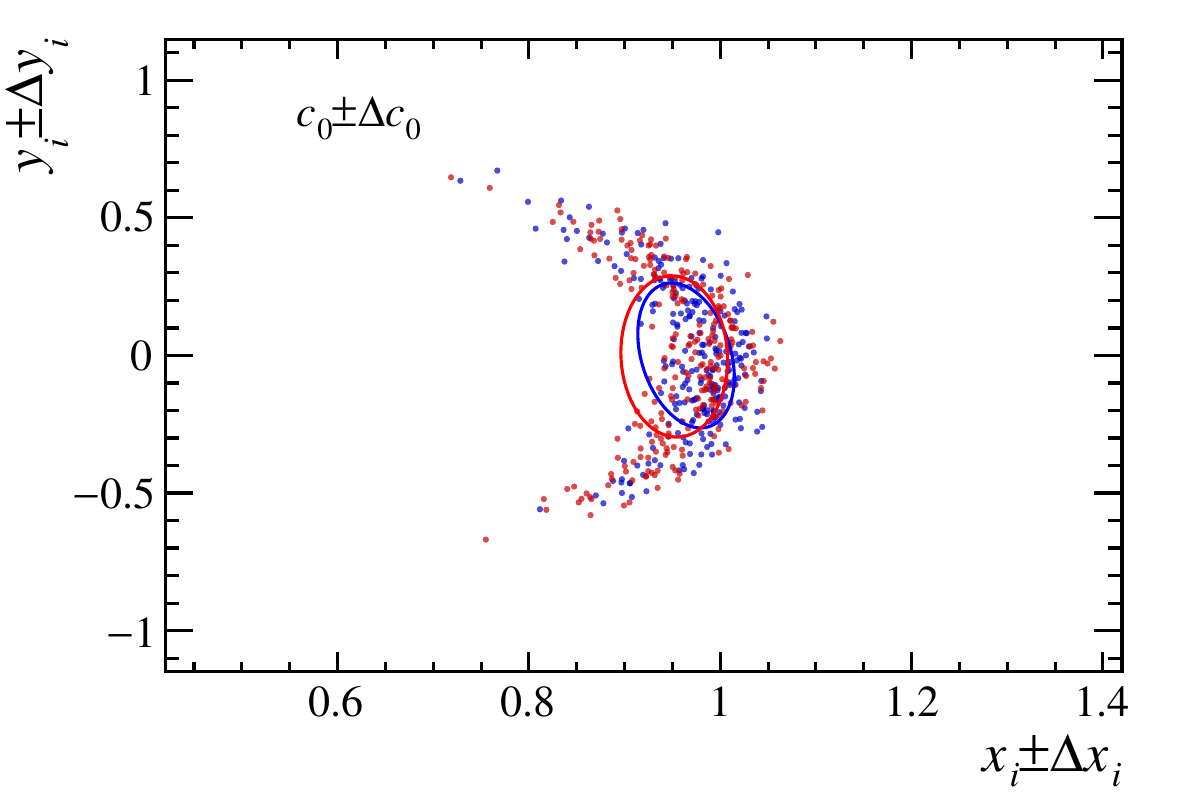}
    \includegraphics[width=0.49\linewidth]{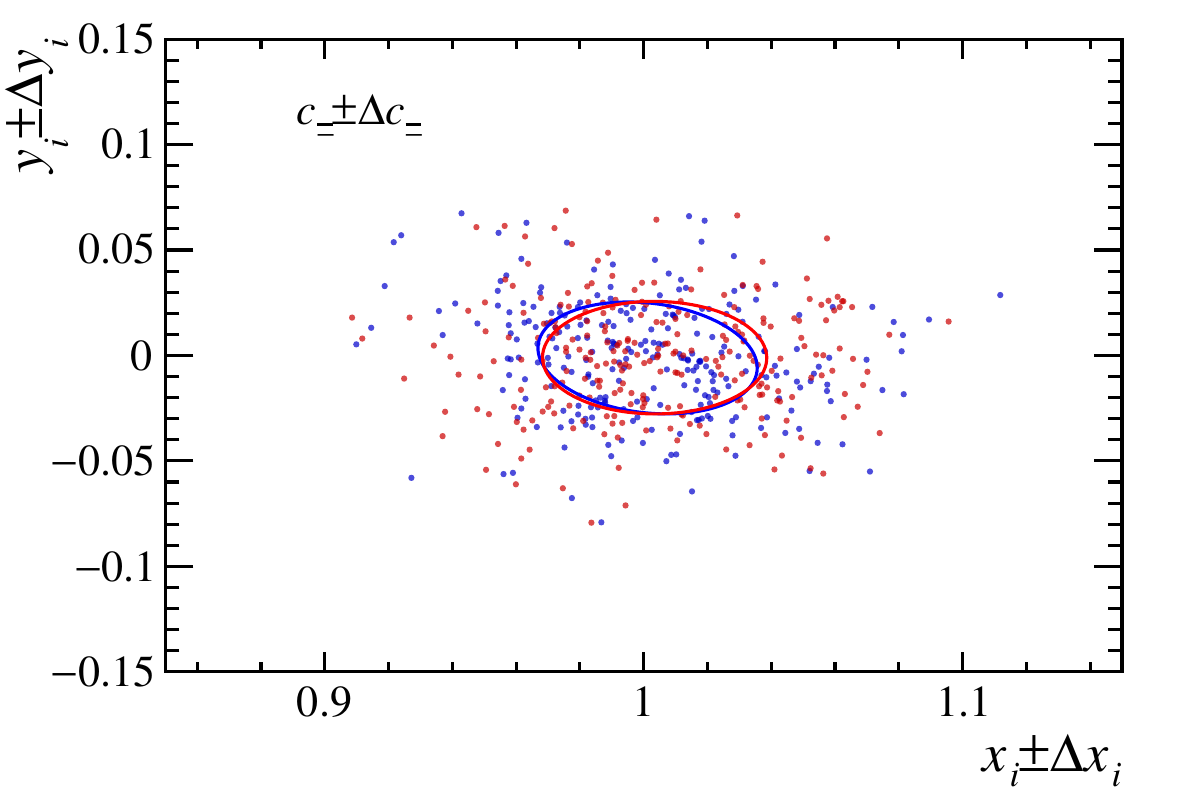}
    \includegraphics[width=0.49\linewidth]{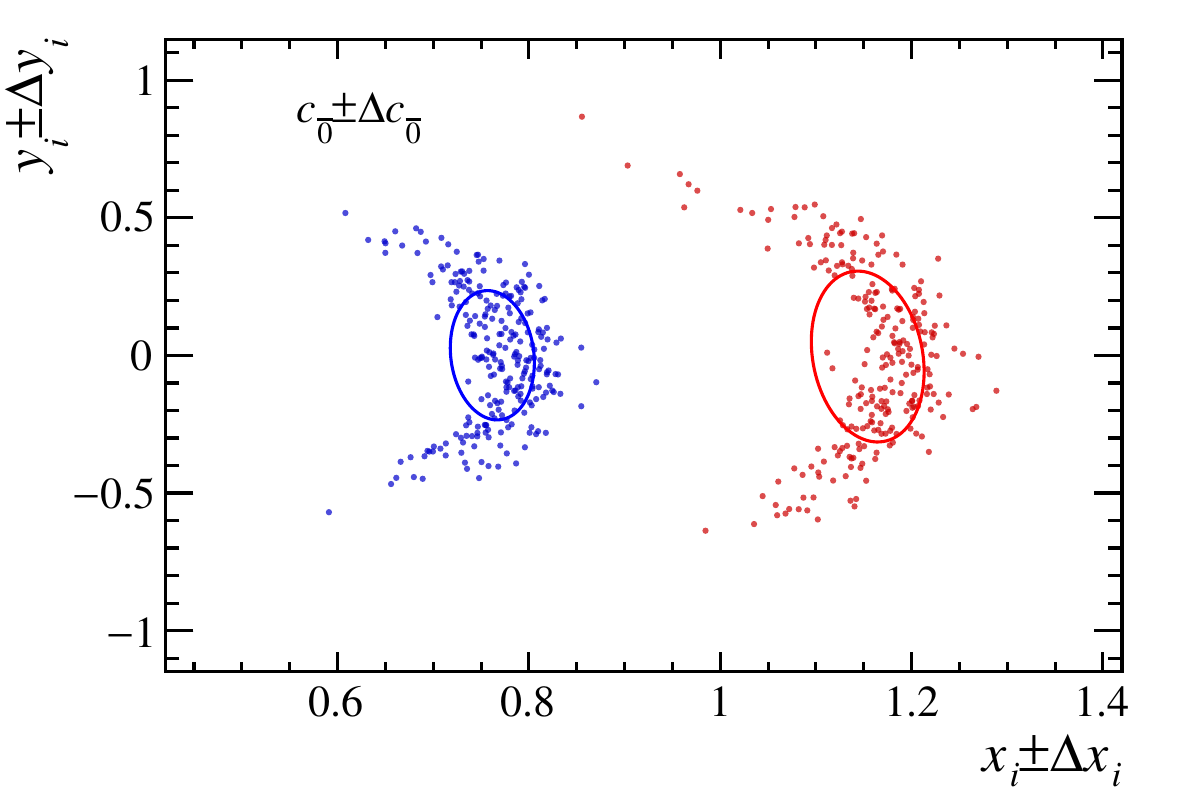}
    \caption{Real and imaginary parts of the amplitude coefficients obtained from fits to pseudoexperiments in Scenario 4~(b), where $\Delta c_{\bar{0}}=-0.2$, as listed in Table~\ref{tab:configs}. Blue and red dots correspond to the amplitude coefficients with plus and minus signs, respectively. The contours indicate the $1\sigma$ regions.}
    \label{fig:scatter}
\end{figure}

\begin{figure}[tbp]
    \centering
    \includegraphics[width=0.325\linewidth]{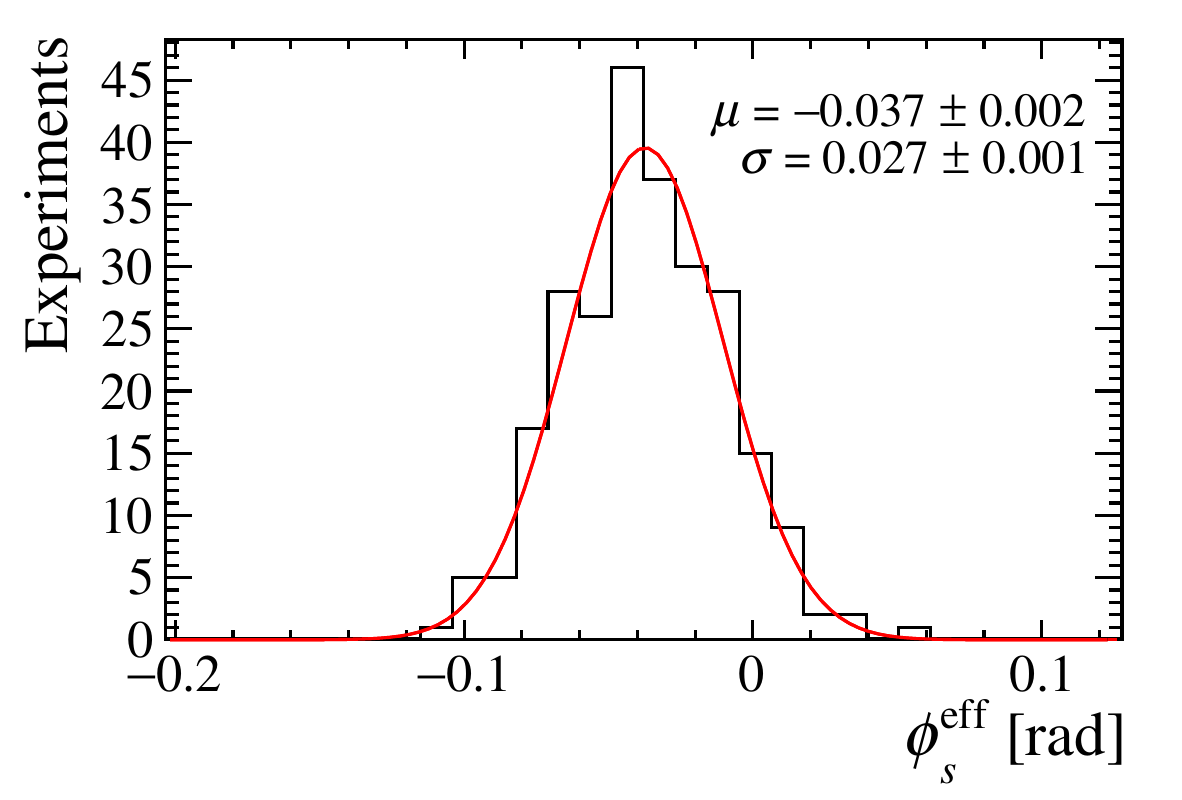}
    \includegraphics[width=0.325\linewidth]{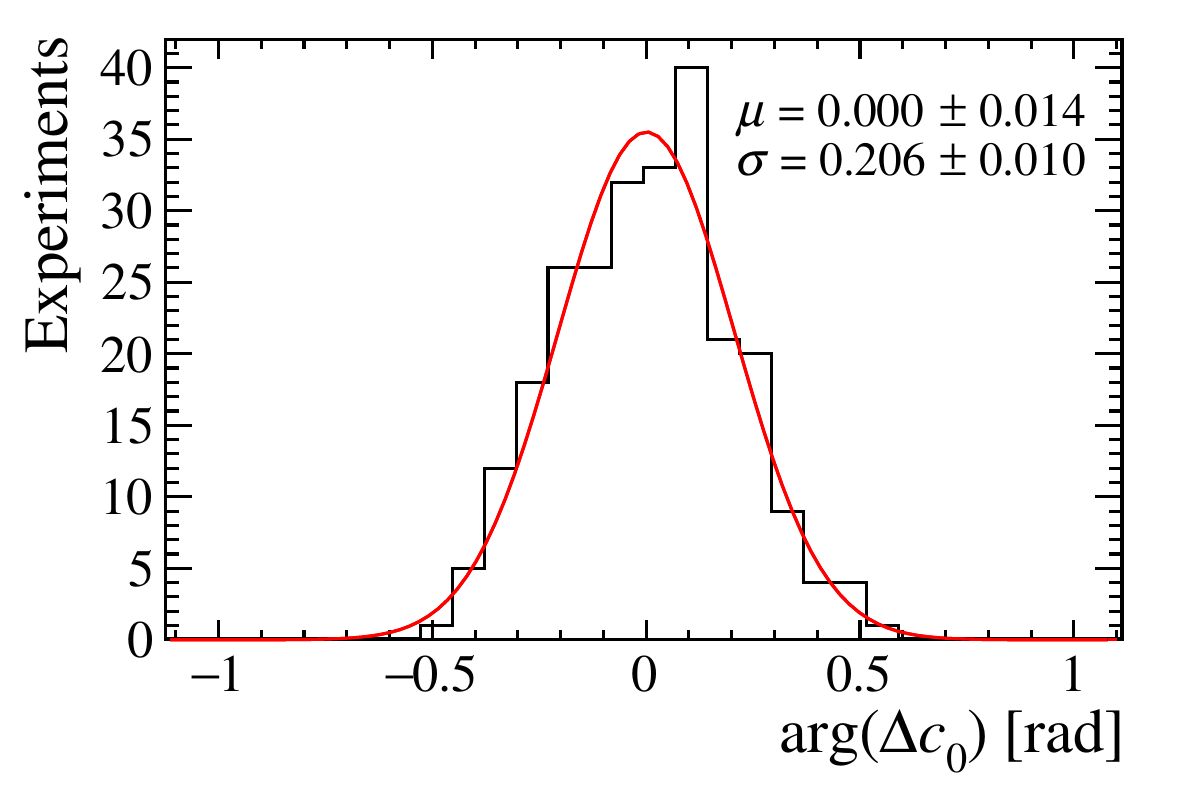}
    \includegraphics[width=0.325\linewidth]{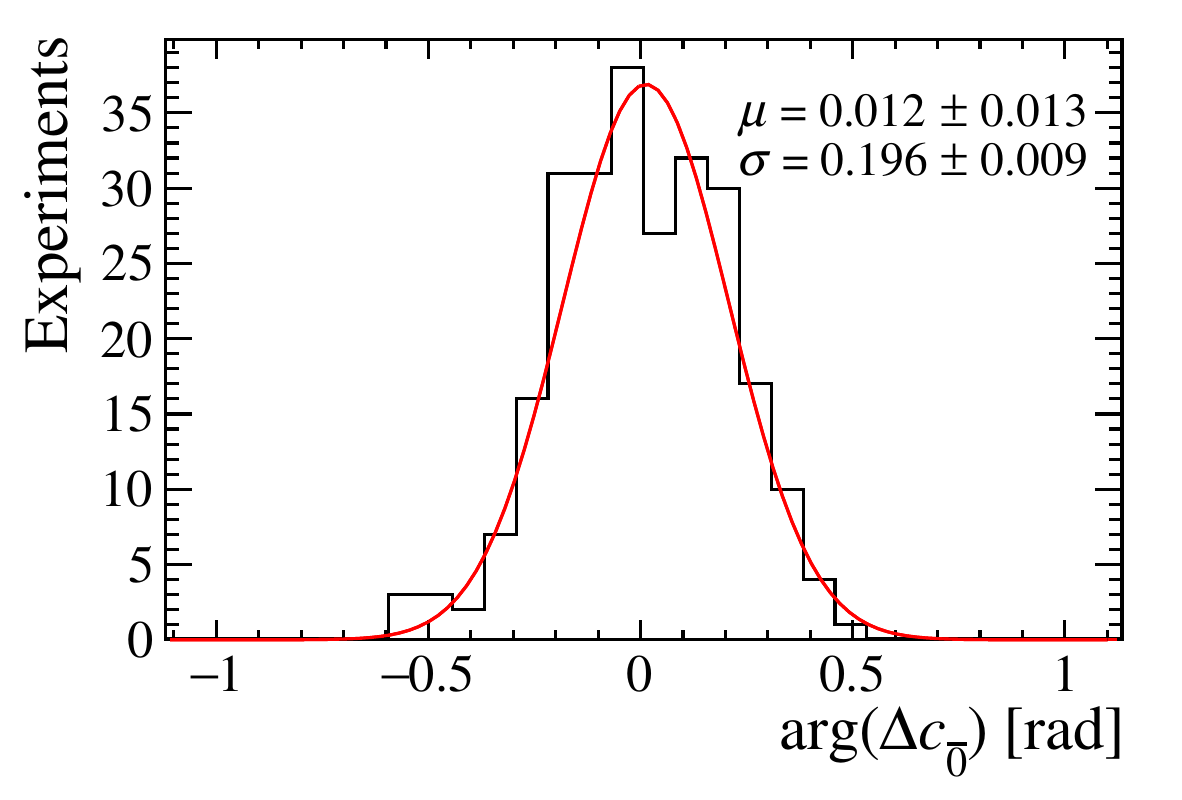}
    \caption{Distributions of $\phi_s^{\rm eff}$ and phases of $\Delta c_0$ and $\Delta c_{\bar{0}}$ determined from fits to pseudoexperiments in Scenario 4~(b), where $\Delta c_{\bar{0}}=-0.2$, as listed in Table~\ref{tab:configs}.}
    \label{fig:phases}
\end{figure}

Fits to the ensembles of pseudoexperiments in all of the scenarios demonstrate that it is possible to determine the input parameters reliably.  
As an example, the fit results for Scenario 4~(b) with $\Delta c_{\bar{0}}=-0.2$ are shown in Figs.~\ref{fig:scatter} and \ref{fig:phases}. 
Fig.~\ref{fig:scatter} shows scatter plots of the fitted values of the real and imaginary parts of the amplitude coefficients determined from fits to the pseudoexperiments. 
The points form arc-like patterns in $c_0 \pm \Delta c_0$ and $c_{\bar{0}} \pm \Delta c_{\bar{0}}$, indicating that the magnitudes are determined more precisely than the phases.
The effect of the non-zero value of $\Delta c_{\bar{0}}$ is also clearly visible.
For the $c_{\bar{-}} \pm \Delta c_{\bar{-}}$ parameters, both real and imaginary parts are more precisely determined due to the overlap in phase space with the reference \Kstarm amplitude, so the arc-like shape is not visible.  

Fig.~\ref{fig:phases} shows the distributions obtained from the fits of the \CP-violating phases $\phi_s^{\rm eff}$, $\arg\left(\Delta c_0\right)$ and $\arg\left(\Delta c_{\bar{0}}\right)$; the phases of $\Delta c_-$ and $\Delta c_{\bar{-}}$ are not shown as these are affected by the constraints. 
The mean of the distribution of fitted $\phi_s^{\rm eff}$ values is consistent with the input value, and the width of this distribution, $26\mrad$, shows good sensitivity to this parameter.
For comparison, the latest $\phi_s$ measurements from LHCb~\cite{LHCb-PAPER-2023-016} and CMS~\cite{CMS:2026qdg} with $\Bs \to \jpsi\phi$ decays each have statistical precision of 22\mrad, while the latest $\phi_s^{\rm eff}$ measurement from LHCb with $\Bs \to\phi\phi$ decays has statistical precision of 75\mrad~\cite{LHCb-PAPER-2023-001}.
It must be stressed that the pseudoexperiments do not account for experimental effects such as background and decay-time resolution, and the model used is a simplistic one, so that the achievable precision from real data will not be as good as obtained here.  

The pseudoexperiments also demonstrate sensitivity to the \CP-violating phases $\arg\left(\Delta c_0\right)$ and $\arg\left(\Delta c_{\bar{0}}\right)$.
The width of the distributions, around 200\mrad, indicates considerably lower sensitivity than that for $\phi_s^{\rm eff}$. 
The difference can be understood from the fact that sensitivity to $\phi_s^{\rm eff}$, with the choice of constraints discussed in the previous section, is dominated by the decay-time-dependent interference between the amplitudes associated with the $c_{\Kstarp}$ and $\bar{c}_{\Kstarp}$ coefficients, and between those associated with $c_{\Kstarm}$ and $\bar{c}_{\Kstarm}$ coefficients, which overlap in phase space and for which a subset of parameters are fixed.
On the other hand, the phases of the amplitudes associated to the \Kstarz and \Kstarzb resonances in both final states have to be determined relative to the reference amplitudes associated with the charged \Kstar resonances, and are therefore limited by the small overlap in phase space. 
This is confirmed by changing constraint~\ref{constraint3} to be associated with the \Kstarzb resonance, while constraints~\ref{constraint1} and~\ref{constraint2} remain unchanged. 
In that case, the reduced overlap between reference amplitudes leads to a larger uncertainty on $\phi_s^{\rm eff}$ than in the case where all reference amplitudes are associated with charged (or neutral) \Kstar resonances. 
This observation indicates that the choice of reference amplitudes can affect the sensitivity to $\phi_s^{\rm eff}$, and an appropriate choice is essential for achieving optimal precision.

\begin{figure}[tbp]
    \centering
    \includegraphics[width=0.49\linewidth]{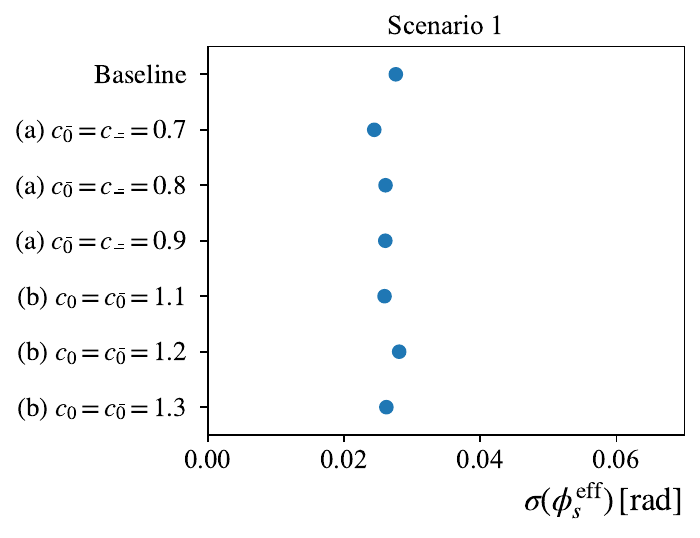}
    \includegraphics[width=0.49\linewidth]{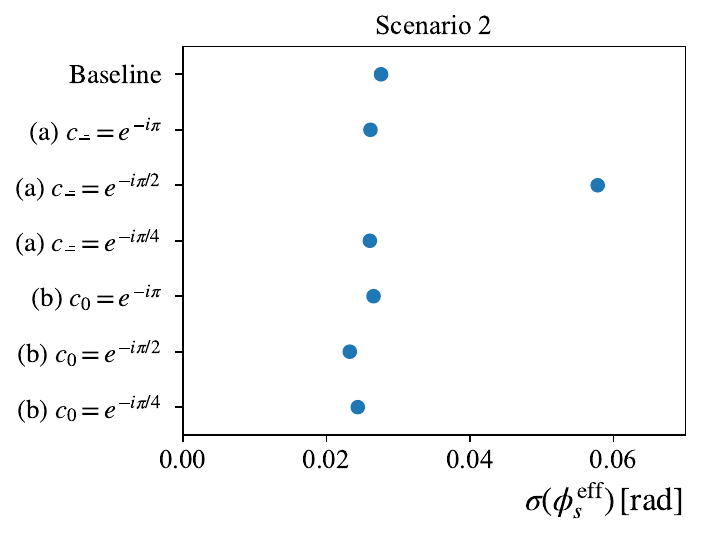}
    \includegraphics[width=0.49\linewidth]{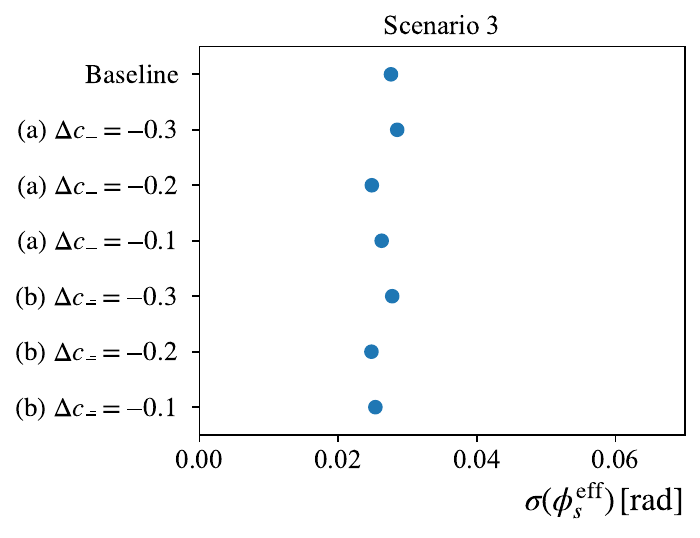}
    \includegraphics[width=0.49\linewidth]{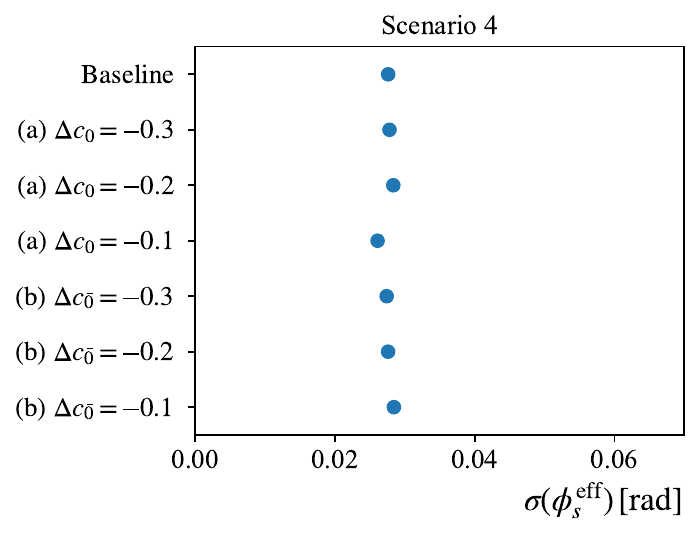}
    \includegraphics[width=0.49\linewidth]{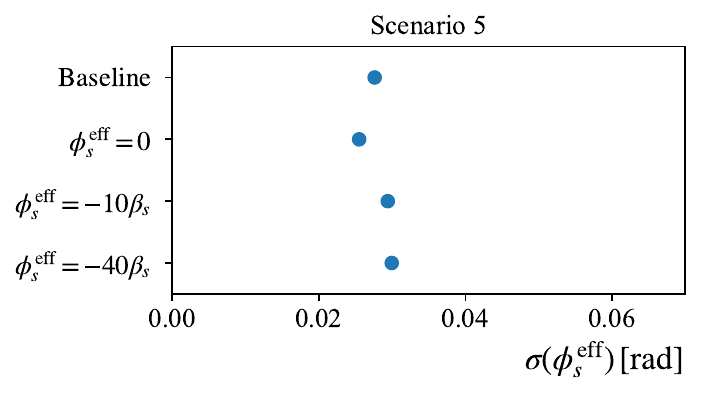}

    \caption{Uncertainties on $\phi_s^{\rm eff}$ for a selected set of model configurations listed in Table~\ref{tab:configs}.}
    \label{fig:phis_comparision}
\end{figure}

The precision on $\phi_s^{\rm eff}$ is compared across a set of representative model configurations, as shown in Fig.~\ref{fig:phis_comparision}. The corresponding uncertainties remain relatively stable.
A larger uncertainty is found in Scenario 2~(a) with $c_{\bar{-}} = e^{-i\pi/2}$. 
To investigate this feature further, more configurations are examined around this point by varying $\arg(c_{\bar{-}})$. 
A consistent pattern is observed. 
The uncertainty increases as $\arg(c_{\bar{-}})$ approaches $-\pi/2$ and decreases as it moves away from that value. 
This is understood as a consequence of the simplistic model used in the pseudoexperiment study.
In the limit that each final state contains only a single resonance, \ie\ the quasi-two-body limit, the decay-time distributions of Eqs.~\eqref{eq:cp_uta:td_cp_bsb_asp1}--\eqref{eq:cp_uta:td_cp_bs_asp1} reduce to the coefficient of the sine terms being given by $\pm \sin(\delta_{f} \pm \phi_s^{\rm  eff})$ and those of the hyperbolic sine terms being given by $-\cos(\delta_{f} \pm \phi_s^{\rm eff})$, where $\delta_{f}$ is the strong phase difference between $\mathcal{\bar{A}}_f$ and $\mathcal{A}_f$, with similar simplifications of Eqs.~\eqref{eq:cp_uta:td_cp_bsb_asp2}--\eqref{eq:cp_uta:td_cp_bs_asp2} except with $f \longrightarrow \bar{f}$.
Since the majority of the precision to $\phi_s^{\rm eff}$ comes from the sine terms, there is intrinsically less sensitivity to small values of $\phi_s^{\rm eff}$ when $\delta_{f}$ or $\delta_{\bar{f}} \approx \pm \frac{\pi}{2}$.
This behaviour indicates that the sensitivity to $\phi_s^{\rm eff}$ depends on the underlying amplitude model.
However, less strong dependence is expected for a more realistic amplitude model including, for example, contributions from $K^*_0(1430)$ states, as the quasi-two-body limit becomes less relevant.

To provide insight into the achievable sensitivity to $\phi_s^{\rm eff}$ with LHCb Run~4 and future Runs~5--6 data, pseudoexperiments with increased statistics are generated and fitted. These samples contain 14\,500 and 80\,000 perfectly tagged decays, corresponding to the expected datasets by the end of Run~4 and Run~6, \ie\ $60\invfb$ and $300\invfb$ of high-energy proton proton collisions, respectively. 
The resulting uncertainties on $\phi_s^{\rm eff}$ are $17\mrad$ and $8.6\mrad$, which follow approximately the expected inverse square-root scaling with the integrated luminosity.
An experimental measurement is required in order to be able to assess whether systematic uncertainties will limit the precision. 

\section{Summary}
\label{sec:Summary}

The \BsToKSKPi decays are of interest to search for beyond-the-SM sources of \CP violation, but require a novel tagged decay-time-dependent Dalitz-plot analysis performed simultaneously in two final states. 
This approach has been explored and demonstrated to be feasible using pseudoexperiments. 
Although the achievable sensitivity depends on the amplitude model and several experimental aspects, an interesting level of precision on $\phi_s^{\rm eff}$ appears possible with the LHCb Runs~1--3 dataset. 
The sensitivity is further projected for the expected LHCb Runs~4--6 data, providing an indication of the achievable precision in the future. 
The method developed in this work can be extended to other tagged decay-time-dependent analyses of multibody decays with two charge-conjugate final states.
The counterpart \Bd\ decay, \ie\ \BdToKSKPi, is of particular interest since the decay amplitudes associated with the neutral \Kstar resonances can be related between the \Bs\ and \Bd\ systems using U-spin, providing control over hadronic uncertainties associated with the SM contributions.  
% which can be pursued at LHCb and Belle~II once sufficient statistics are available.

\section*{Acknowledgements}
The authors wish to thank their colleagues on the LHCb experiment for the fruitful and enjoyable collaboration. 
In particular, they would like to thank Fr\'{e}d\'{e}ric Blanc for helpful comments.
This project has received funding from the European Union’s Marie Skłodowska-Curie Actions (MSCA) under grant agreement No.~101207117, for the project \texttt{SimDalitzCPV}.
TG and TL are supported by the Science and Technology Facilities Council (UK).

\clearpage

\addcontentsline{toc}{section}{References}
%\setboolean{inbibliography}{true}
\bibliographystyle{LHCb}
\bibliography{main,standard,LHCb-PAPER,LHCb-DP}
 
\end{document}